\documentstyle[sprocl,epsfig]{article}

\def\be{\begin{equation}}
\def\ee{\end{equation}}
\def\bea{\begin{eqnarray}}
\def\eea{\end{eqnarray}}

\begin{document}

\vspace*{-9\baselineskip}
\begin{flushright}
\hfill {\rm UR-1597}\\
\hfill {\rm ER/40685/942}\\
\hfill {\rm September 1999}\\
\end{flushright}
\vspace*{2\baselineskip}

\title{A MONTE CARLO FOR BFKL PHYSICS\footnote{Presented at the 
International Workshop on Linear Colliders, Sitges, Spain, April 28-May 5, 1999}}

\author{ LYNNE H.~ORR }

\address{Department of Physics and Astronomy, University of Rochester\\
Rochester, NY 60627-0171, USA}

\author{ W.J.~STIRLING }

\address{Departments of Physics and Mathematical Sciences, University of
Durham\\ Durham DH1 3LE, UK}

%%%%%%%%%%%%%%%%%%%%%%%%%%%%%%%%%%%%%%%%%%%%%%%%%%%%%%%%%%%%%%
% You may repeat \author \address as often as necessary      %
%%%%%%%%%%%%%%%%%%%%%%%%%%%%%%%%%%%%%%%%%%%%%%%%%%%%%%%%%%%%%%

\maketitle\abstracts{
Virtual photon scattering in $e^+e^-$ collisions can 
result in events with the electron-positron pair at large
rapidity separation with hadronic activity in between.  
The BFKL equation resums large logarithms that  dominate
the cross section for this process.    We report here
on a Monte Carlo method for solving the BFKL equation that 
allows kinematic constraints to be taken into account.  
The application to $e^+e^-$ collisions is in progress.
}
  
\section{Introduction}

High energy $e^+e^-$ collisions can lead to the scattering
of virtual photons emitted by the initial electron and positron.  
When the virtuality  $Q^2$ of these photons is small compared to
the center-of-mass energy $\hat{s}$ of the $\gamma^*\gamma^*$
system, the scattering cross section is dominated by 
contributions in which the photons split into quark-antiquark
pairs, with t-channel gluon exchange.  The emission of additional
soft gluons from the t-channel gluon gives rise to large logarithms that 
lead to corrections in powers of 
\be
\alpha_s(Q^2)\ln(\hat{s}/Q^2),
\ee
which is of order one in this kinematic regime.  These logarithms
must therefore be resummed in the calculation of the cross section.
The  events that result from this process are characterized by 
electron-positron pairs with a large rapidity separation, and hadronic
activity in between. 

The large-logarithm resummation is performed by the 
Balitsky-Fadin-Kuraev-Lipatov (BFKL) equation~\cite{bfkl},
where its analytic solution gives a rise in the cross section
$\hat\sigma \sim (\hat{s})^\lambda$,
with  $\lambda = 4C_A\ln 2\, \alpha_s/\pi \approx 0.5$.
The BFKL equation applies not only to virtual photon scattering as
described above, but also to dijet production at 
large rapidity difference in hadron-hadron collisions and 
to forward jet production in lepton-hadron collisions. 

The BFKL equation can be solved analytically, but to do so requires giving up 
energy-momentum conservation, because it involves integration
over arbitrarily large momenta of emitted gluons.  Furthermore, because
the sum over gluons is implicit, only leading-order kinematics can be included.
This leads to predictions that do not correspond to any real experimental
situation.  In principle the corrections due to kinematic effects are 
in subleading, but in practice, as we will see below, they can 
be quite important.

\section{A Monte Carlo for BFKL Physics}

The solution to the problem of lack of kinematic constraints in analytic
BFKL predictions is to unfold the implicit gluon sum to make it
explicit, and to implement the result in a Monte Carlo
event generator~\cite{os,schmidt}.  This is achieved as follows.
The BFKL equation contains separate integrals over  real and virtual 
emitted gluons.  We  combine the 
`unresolved' real emissions --- those with transverse momenta
below some minimum value (small
compared to the momentum threshold for measured
jets) --- with the virtual emissions.  Schematically,
we have 
\be
\int_{virtual} + \int_{real} = \int_{virtual+real, unres.} +
\int_{real, res.}
\ee
We  perform
the integration over virtual and unresolved real
emissions  analytically.  

We  then solve the  
BFKL equation
by iteration, and we obtain a differential cross section
that contains an explicit sum over emitted gluons along with 
the appropriate phase space factors.  In addition, we obtain
an overall form factor due
to virtual and unresolved emissions. The subprocess cross section is
\be
d\hat\sigma=d\hat\sigma_0\times\sum_{n\ge 0} f_{n}
\ee
where $f_{n}$ is the iterated solution for $n$ real gluons emitted and
contains the overall form factor.
It is then straightforward to implement the result in a Monte Carlo
event generator.   Emitted real (resolved) gluons appear explicitly, 
so that conservation of momentum and energy 
is based on exact kinematics
for each event.  In addition, we include the running of the strong
coupling constant.  See~\cite{os} for further details.

\section{Results and Prospects}

\begin{figure*}[t]
\psfig{figure=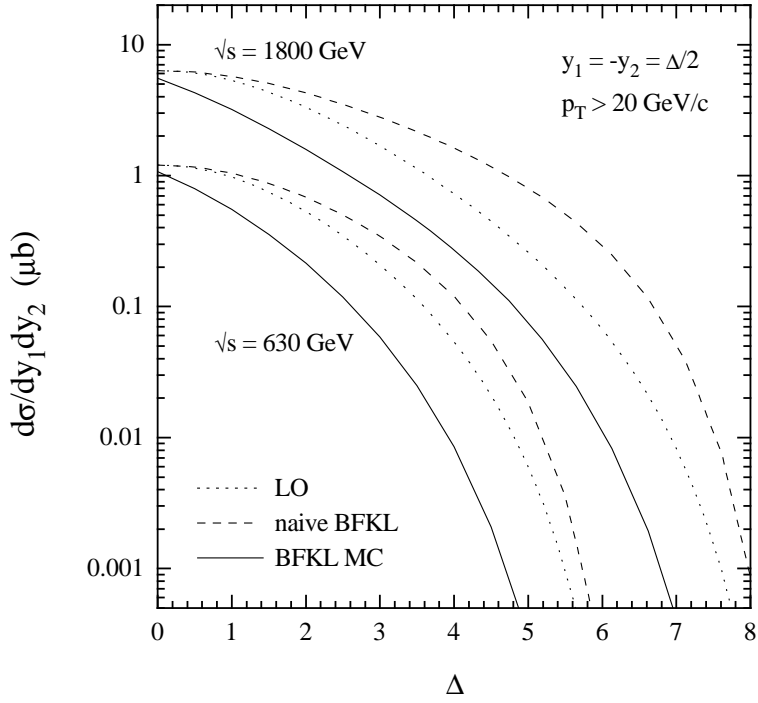,width=11cm}
\vskip -.75cm
\caption[]{  The dependence on the dijet rapidity separation
of the BFKL and asymptotic QCD leading-order 
dijet cross sections  at the Tevatron~\cite{osbis}.
The three curves at each collider energy
 use: (i) `improved' BFKL MC (solid lines), (ii) `naive' BFKL (dashed lines),
  and (iii) the asymptotic ($\Delta \gg 1$) form
 of QCD leading order (dotted lines).\label{fig:dsigma}}
\end{figure*}

We have used this BFKL Monte Carlo approach to study dijet production at
hadron colliders in detail~\cite{os,osbis,oslhc}.  
The most important conclusion is that the effects of kinematic constraints can 
be {\it very} large,  because they suppress radiation of the gluons that give 
rise to what are considered to be characteristic BFKL effects.  As
a result, the predictions can change substantially.  This is illustrated in
Fig.~\ref{fig:dsigma}, which shows the dijet cross section at the Tevatron
for two different center-of-mass energies as a function of the dijet rapidity
difference.  The naive analytic BFKL prediction lies above the leading
QCD curve, as expected.  But when kinematic constraints are included, the 
BFKL prediction gets pushed {\it below} that of leading-order QCD.
Clearly it is important to incorporate kinematic constraints in 
our BFKL predictions.

We are currently completing the application of our BFKL Monte Carlo to 
virtual photon scattering in $e^+e^-$ collisions and in forward
jet production at HERA.  In both cases we expect kinematic
constraints to be large and to lead to some suppression of 
BFKL effects.

\section*{Acknowledgments}
Work supported in part by the U.S. Department of Energy,
under grant DE-FG02-91ER40685 and by the U.S. National Science Foundation, 
under grant PHY-9600155.

\end{document}